\documentclass[twoside,twocolumn,9pt]{article}
\usepackage{extsizes}
\usepackage[super,sort&compress,comma]{natbib} 
\usepackage[version=3]{mhchem}
\usepackage[left=1.5cm, right=1.5cm, top=1.785cm, bottom=2.0cm]{geometry}
\usepackage{units}
\usepackage{balance}
\usepackage{widetext}
\usepackage[normalem]{ulem}
\usepackage{times,mathptmx}
\usepackage{sectsty}
\usepackage{graphicx} 
\usepackage{lastpage}
\usepackage[format=plain,justification=raggedright,singlelinecheck=false,font={stretch=1.125,small,sf},labelfont=bf,labelsep=space]{caption}
\usepackage{float}
\usepackage{fancyhdr}
\usepackage{fnpos}
\usepackage[english]{babel}
\usepackage{array}

\usepackage{droidsans}
\usepackage{charter}
\usepackage[T1]{fontenc}
\usepackage[usenames,dvipsnames]{xcolor}
\usepackage{setspace}
\usepackage[compact]{titlesec}

\usepackage{xspace}

\definecolor{cream}{RGB}{222,217,201}

\begin{document}

\pagestyle{fancy}
\thispagestyle{plain}
\fancypagestyle{plain}{

\fancyhead[C]{\includegraphics[width=18.5cm]{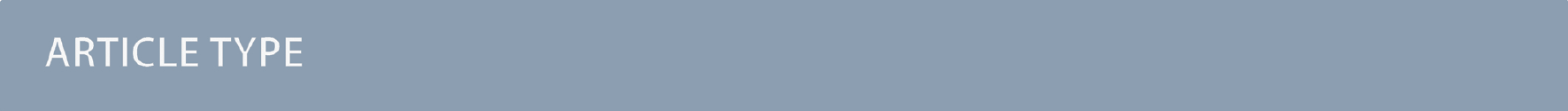}}
\fancyhead[L]{\hspace{0cm}\vspace{1.5cm}\includegraphics[height=30pt]{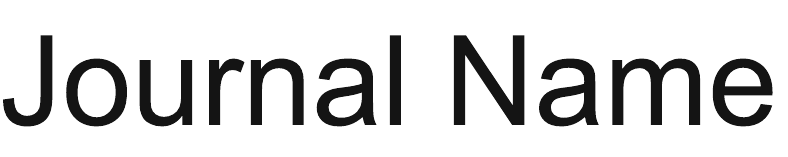}}
\fancyhead[R]{\hspace{0cm}\vspace{1.7cm}\includegraphics[height=55pt]{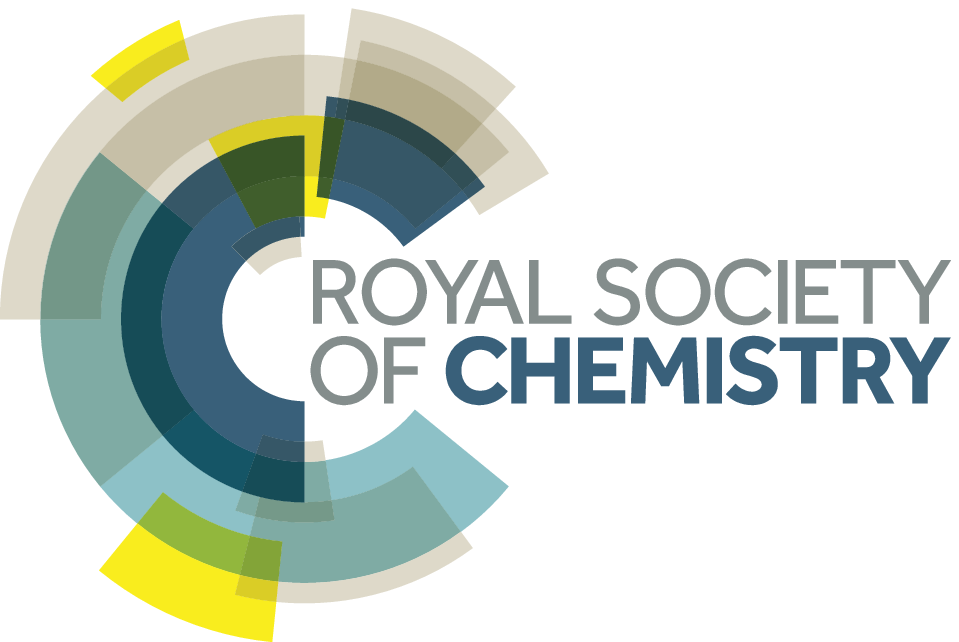}}
\renewcommand{\headrulewidth}{0pt}
}

\makeFNbottom
\makeatletter
\renewcommand\LARGE{\@setfontsize\LARGE{15pt}{17}}
\renewcommand\Large{\@setfontsize\Large{12pt}{14}}
\renewcommand\large{\@setfontsize\large{10pt}{12}}
\renewcommand\footnotesize{\@setfontsize\footnotesize{7pt}{10}}
\makeatother

\renewcommand{\thefootnote}{\fnsymbol{footnote}}
\renewcommand\footnoterule{\vspace*{1pt}%
\color{cream}\hrule width 3.5in height 0.4pt \color{black}\vspace*{5pt}} 
\setcounter{secnumdepth}{5}

\makeatletter 
\renewcommand\@biblabel[1]{#1}            
\renewcommand\@makefntext[1]%
{\noindent\makebox[0pt][r]{\@thefnmark\,}#1}
\makeatother 
\renewcommand{\figurename}{\small{Fig.}~}
\sectionfont{\sffamily\Large}
\subsectionfont{\normalsize}
\subsubsectionfont{\bf}
\setstretch{1.125} 
\setlength{\skip\footins}{0.8cm}
\setlength{\footnotesep}{0.25cm}
\setlength{\jot}{10pt}
\titlespacing*{\section}{0pt}{4pt}{4pt}
\titlespacing*{\subsection}{0pt}{15pt}{1pt}

\fancyfoot{}
\fancyfoot[LO,RE]{\vspace{-7.1pt}\includegraphics[height=9pt]{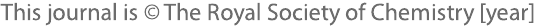}}
\fancyfoot[CO]{\vspace{-7.1pt}\hspace{13.2cm}\includegraphics{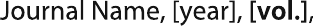}}
\fancyfoot[CE]{\vspace{-7.2pt}\hspace{-14.2cm}\includegraphics{head_foot/RF}}
\fancyfoot[RO]{\footnotesize{\sffamily{1--\pageref{LastPage} ~\textbar  \hspace{2pt}\thepage}}}
\fancyfoot[LE]{\footnotesize{\sffamily{\thepage~\textbar\hspace{3.45cm} 1--\pageref{LastPage}}}}
\fancyhead{}
\renewcommand{\headrulewidth}{0pt} 
\renewcommand{\footrulewidth}{0pt}
\setlength{\arrayrulewidth}{1pt}
\setlength{\columnsep}{6.5mm}
\setlength\bibsep{1pt}

\makeatletter 
\newlength{\figrulesep} 
\setlength{\figrulesep}{0.5\textfloatsep} 

\newcommand{\topfigrule}{\vspace*{-1pt}%
\noindent{\color{cream}\rule[-\figrulesep]{\columnwidth}{1.5pt}} }

\newcommand{\botfigrule}{\vspace*{-2pt}%
\noindent{\color{cream}\rule[\figrulesep]{\columnwidth}{1.5pt}} }

\newcommand{\dblfigrule}{\vspace*{-1pt}%
\noindent{\color{cream}\rule[-\figrulesep]{\textwidth}{1.5pt}} }

\newcommand{\bjoern}[2]{\sout{#1}\textcolor{green}{#2}}

\newcommand{\equ}[1]{eq.~\ref{equ:#1}}
\newcommand{\tab}[1]{Tab.~\ref{tab:#1}}
\newcommand{\Equ}[1]{Eq.~\ref{equ:#1}}
\newcommand{\fig}[1]{\ref{fig:#1}}
\newcommand{\sect}[1]{sec.~\ref{sec:#1}}
\newcommand{\Fig}[1]{Figure~\ref{fig:#1}}
\newcommand{\ppes}{poly {\em para} phenylene ethynylenes\xspace}
\newcommand{\ppe}{poly {\em para} phenylene ethynylene\xspace}

\makeatother

\twocolumn[
  \begin{@twocolumnfalse}
\vspace{3cm}
\sffamily
\begin{tabular}{m{4.5cm} p{13.5cm} }

\includegraphics{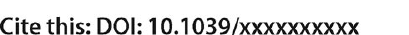} & \noindent\LARGE{\textbf{Getting excited: Challenges in quantum-classical studies of excitons in polymeric systems $^\dag$}} \\
\vspace{0.3cm} & \vspace{0.3cm} \\

& \noindent\large{Behnaz Bagheri,\textit{$^{a}$} Bj\"orn Baumeier,\textit{$^{a \ast}$} and Mikko Karttunen\textit{$^{a \ast}$}} \\

\includegraphics{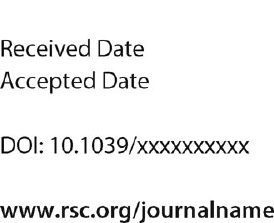} & \noindent\normalsize{A combination of classical molecular dynamics (MM/MD) and quantum chemical calculations based on the density functional theory (DFT) was performed to describe conformational  properties of diphenylethyne (DPE), methylated-DPE and poly para phenylene ethynylene (PPE). DFT calculations were employed to improve and develop force field parameters for MM/MD simulations. Many-body 
Green's functions theory within the GW approximation and the Bethe-Salpeter equation were utilized to describe excited states of the systems. Reliability of the excitation energies based on the MM/MD  conformations was examined and compared to the excitation energies from DFT conformations. The results show an overall agreement between the optical excitations based on MM/MD conformations and DFT conformations. This allows for calculation of excitation energies based on MM/MD conformations. 
}
\end{tabular}

 \end{@twocolumnfalse} \vspace{0.6cm}

  ]

\renewcommand*\rmdefault{bch}\normalfont\upshape
\rmfamily
\section*{}
\vspace{-1cm}


\footnotetext{\textit{$^{a}$~Department of Mathematics and Computer Science \& Institute for Complex Molecular Systems, Eindhoven University of Technology, P.O. Box 513, 5600 MB Eindhoven, The Netherlands. }}
\footnotetext{\textit{$^{\ast}$ E-mail: b.baumeier@tue.nl, m.e.j.karttunen@tue.nl }}

\footnotetext{\dag~Electronic Supplementary Information (ESI) available: [details of any supplementary information available should be included here]. See DOI: 10.1039/b000000x/}




\section{Introduction}
\label{sec:intro}
Multiscale modelling has become one the leading themes in modelling materials and their different properties. The most famous use of the term is the 2013 Nobel Prize in Chemistry "for the development of multiscale models for complex chemical systems". Its increasing popularity can also been seen in the titles of published peer reviewed papers: According to the Web of Science, in 2015 1,157 articles had the term "multiscale" in their title, a decade earlier the number was 520 and in 1990 only 41. 

"Multiscale" is often used synonymously with "coarse graining".  
Coarse graining typically refers to obtaining interaction potentials
and parameters for a higher level system from structural equilibrium data. Examples of such are force matching~\cite{Izvekov2005,Lu2010}, inverse Boltzmann~\cite{Reith03} and the inverse Monte Carlo method~\cite{Lyubartsev1995,Lyubartsev2002}. The latter two are based on the Henderson theorem~\cite{Henderson1974a} which is essentially the Hohenberg-Kohn theorem~\cite{Hohenberg:64ht} for classical systems; for a brief derivation and discussion of the relation between them, see Murtola et al.~\cite{Murtola2009d}. More heuristic approaches such the Martini model~\cite{Marrink2013} and the PLUM model~\cite{Bereau2014} are another common approach; for a comparison between PLUM, Martini and atomistic models, see Bereau et al.~\cite{Bereau2015}. One of the leading ideas is that systematic coarse graining allows, at least in principle, also fine graining, that is, remapping the higher level model back to the original more microscopic one~\cite{Krishna2011b}. Backmapping procedures exist for the Martini model~\cite{Rzepiela2010a,Wassenaar2014} and such approaches have proven useful in modelling lipids and proteins, see e.g. Pannuzzo et al.~\cite{Pannuzzo2013a}. One somewhat less considered but an important issue is sampling at differerent levels of coarse-graining~\cite{Lyman2006}. 
For more details on coarse graining, recent reviews can be found in Refs.~\citenum{Murtola2009d,Ingolfsson2013,Noid2013,Brini2013}

Multiscale modelling is a much broader concept. For example, instead of linking scales via deriving new interaction potentials, in hybrid simulations some part of the system is described with a different resolution from the rest and information is transmitted between the two different regions. Examples of such are QM/MM (quantum-molecular mechanics), MM/CG (molecular mechanics-coarse grained) and even QM/QM (quantum-quantum)). The idea  is that more detailed information in some well defined region is sought after and the crucial issue is how to couple the main system and the subsystem. This has been discussed extensively, see, e.g. Refs.~\citenum{Pezeshki2014}, but the essence is that the both dynamic and static properties must be communicated between the systems. This may include polarization, changes in the charge-state of the system, and so on. 
Yet another multiscale approach is the so-called adaptive resolution method, or AdResS. In this case, run-time information is transmitted  between layers of description ranging from atomistic even up to to continuum~\cite{Praprotnik2008c,Delgado-Buscalioni2009b}. 

Electronic excitations pose a significant challenge for multiscaling~\cite{olivier_theoretical_2009,di_donato_n-type_2010,nelson_modeling_2009,vehoff_charge_2010,vehoff_charge_2010-1,ruehle_multiscale_2010,wang_computational_2010,schrader_comparative_2012,may_design_2012} since typical DFT methods describe the ground state. 
An assessment of the interplay between molecular electronic structure, morphological order, and thermodynamic properties requires the knowledge of the material morphology at atomic resolution, as well as strategies to couple quantum mechanical techniques to classical environments for accurate evaluation of electronic excitations~\cite{may_can_2012,baumeier_electronic_2014,poelking_impact_2015}.

Morphology can have several characteristic length scales and be kinetically arrested.
Employing all-atom molecular dynamics is limited to a few microseconds, which might be too short to fully relax molecular positions and orientations during aggregation~\cite{clancy_application_2011} in polymer melts~\cite{zhang_equilibration_2014,potestio_computer_2014,gemunden_nematic_2013}, or polymers in miscible solutions~\cite{mukherji_polymer_2014}. In such cases more coarse representations might be helpful to overcome the limitations of atomistic models~\cite{ruehle_versatile_2009,huang_coarse-grained_2010,moreira_direct_2015,gemunden_effect_2015}.
Using empirical atomistic potentials in multiscale simulations of excitations based on quantum calculations requires that the structural description at different levels of resolution are compatible with each other. 
For example, bond length deviations or fluctuations in angles and torsions can lead to substantial artifacts if the backmapped/fine-grained geometries do not match the potential energy surfaces (PES) of the underlying quantum mechanical system. 
Such a situation regularly arises for conjugated polymers since conjugation can depend sensitively on conformation. In (semi)flexible polymers, conjugation along a single chain can be broken due to large out-of-plane torsions between two repeat units. Broken conjugation and wave function localization~\cite{vukmirovi_charge_2008,vukmirovi_charge_2009,mcmahon_ad_2009} are often intuitively interpreted based on a simple empirical criterion, the dihedral angle between two adjacent repeat units~\cite{ruehle_multiscale_2010}. In general, details are specific to the backbone chemistry, functionalization by side chains, and solute-solvent interactions. Characteristics of conjugation also directly influence the localization behavior of electronic excitations and hence the electronic and optical properties of the polymer.

\begin{figure}
\centering
\includegraphics[width=\linewidth]{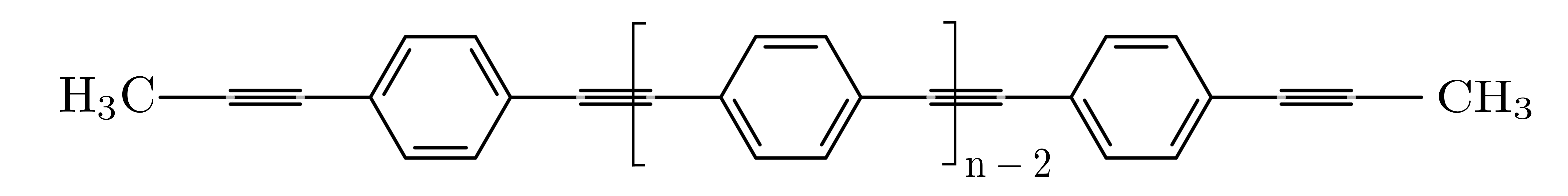}
\caption{Chemical structure of poly {\em para} phenylene ethynylene (poly-PPE). $n$ is the number of repeat units along the polymer (degree of polymerization).}
\label{PPE-chem}       
\end{figure}

\begin{figure}
\centering
\resizebox{0.50\columnwidth}{!}{
\includegraphics{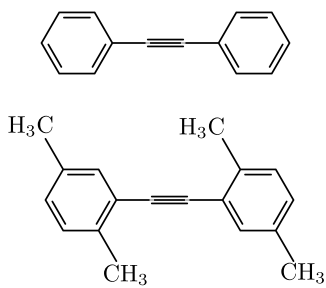}}
\caption{Chemical structures of diphenylethyne (DPE, top) and methylated diphenylethyne (Me-DPE, bottom). It consists of two aromatic rings bridged by a sequence of single bonds and very stiff triple bonds.} 
\label{DPEstructure}     
\end{figure}

In this paper, some of the underlying challenges pertaining to the transfer of structural atomistic detail between quantum and all-atom resolutions are demonstrated. As an example, we consider the calculation of optical properties of \ppe (poly-PPE, see chemical structure in Fig.~\ref{PPE-chem}), a relatively rigid conjugated polymer consisting of aromatic phenyl rings bridged by alternating single and triple carbon bonds. PPEs can be prepared in a variety of morphologies, ranging from extended single chains to polydots
and their optical properties make them particularly attractive for use in fluorescence imaging and sensing~\cite{wu_multicolor_2008,halkyard_evidence_1998,tuncel_conjugated_2010}. 
Due to the importance of backbone torsions on conjugation and hence excitations, we compare PESs for phenyl rotations in diphenylethyne (DPE, see Fig.~\ref{DPEstructure}) obtained using density functional theory (DFT) to the ones from all-atom simulations using the standard force field and experimental data. 
Significant discrepancies were found and 
as a result, the atomistic force field was re-parameterized. 
With this modified force field, ground state geometries are optimized for $n$-PPE oligomers with $n=1,\dots,10$ and then used in $GW$-BSE calculations. The associated excitation energies are benchmarked with results from quantum-mechanical treatment, revealing qualitatively similar characteristics as a function of $n$ but deviations at the quantitative level. Finally, conformations from MD simulations of 2,5-dinonyl-10-PPE solvated in toluene are used in a QM/MM setup to evaluate absorption and emission spectra.

\section{Methodology}
\label{sec:methodology}

MM/MD calculations were performed using a force field of OPLS (optimized potential for liquid simulations)~\cite{jorgensen_optimized_1984,jorgensen_development_1996,watkins_perfluoroalkanes:_2001} form with GROMACS simulation software version 4~\cite{van_der_spoel_gromacs:_2005}. The force field parameters are taken from the polymer consistent force field~\cite{sun_ab_1994,Sun1995} (PCFF) as converted to OPLS form in Refs.~\citenum{maskey_internal_2013,maskey_conformational_2011}. We refer to it as PCFF* from now on. 
 The OPLS potential energy function consists of harmonic bond stretching ($V_{\textnormal{bond}}$), angle bending potential ($V_{\textnormal{angle}}$), non-bonded terms ($V_{\textnormal{non-bonded}}$) including Lennard-Jones (LJ) and electrostatics, proper and improper dihedral potential terms ($V_{\textnormal{torsion}}$)~\cite{jorgensen_optimized_1984,jorgensen_development_1996,watkins_perfluoroalkanes:_2001}:
 \begin{equation}
V_{\textnormal{bond}}=\sum_{i}k_{\textnormal{b},i}(r_i-r_{0,i})^{2}
\label{bond}
\end{equation}
 \begin{equation}
V_{\textnormal{angle}}=\sum_{i}k_{\theta,i}(\theta_{i}-\theta_{0,i})^{2}
\label{angle}
\end{equation}
  \begin{equation}
V_{\textnormal{non-bonded}}=\sum_{i}\sum_{j>i}\Big{\{} \frac{q_{i}q_{j}e^2}{r_{ij}} + 4 \epsilon_{ij} \Big{[} \Big{(} \frac{\sigma_{ij}}{r_{ij}} \Big{)}^{12} -  \Big{(}\frac{\sigma_{ij}}{r_{ij}} \Big{)}^{6}  \Big{]} \Big{\}}
\label{non-bonded}
\end{equation}
 \begin{eqnarray}
V_{\textnormal{torsion}}&=&\sum_{i}\Big{[} \frac{1}{2}k_{1,i}(1+\cos(\phi_{i}))+\frac{1}{2}k_{2,i}(1-\cos(2\phi_i))\nonumber\\
&+&\frac{1}{2}k_{3,i}(1+\cos(3\phi_i))+\frac{1}{2}k_{4,i}(1-\cos(4\phi_i))\Big{]}
\label{torsion}
\end{eqnarray}
The parameters $k_{\textnormal{b},i}$ and $k_{\theta,i}$ are the bond force constant for bond $i$ and angle force constant for angle $i$, respectively. $r_{0}$ and $\theta_{0}$ are initial (reference,equilibrium) bond distance and angle bending, respectively. $k_{1,i},k_{2,i},\dots$ are the torsional force constants for each dihedral $i$. $q_ie$ is the partial atomic charge of atom $i$ in which $e$ is the charge of one electron, $\sigma_{ij}$ are the LJ radii and $\epsilon_{ij}$ are the LJ energies (well-depth) and $r_{ij}$ are the distances  between atom $i$ and $j$. The geometric combination rules were used following the convention adopted in OPLS force field [$\sigma_{ij}=(\sigma_{ii}\sigma_{jj})^{\frac{1}{2}}$ and $\epsilon_{ij}=(\epsilon_{ii}\epsilon_{jj})^{\frac{1}{2}}$]. The intramolecular non-bonded interactions were evaluated for atom pairs separated by three or more bonds. The 1,4-intramolecular interactions were reduced~\cite{jorgensen_optimized_1984,jorgensen_development_1996,watkins_perfluoroalkanes:_2001} by a factor of $1/2$.

To obtain relaxed scans of potential energy surfaces (PES) from MM/MD, energy minimization of the DPE molecule in vacuum was performed using  the conjugate gradient method  followed by a short MD run (100\,ps) with constant particle number ($N$) and temperature ($T$). The Langevin thermostat~\cite{grest_molecular_1986} with $1$\,fs time step and open boundary conditions were applied. Temperature was kept at \unit[10]{K} with \unit[10]{fs} damping constant. All LJ interactions were cut-off at \unit[1.2]{nm}. A plain cut-off scheme was used for electrostatic interactions with \unit[2.0]{nm} real space cut-off: with open boundary conditions plain cut-off can be used. For systems with periodic boundary conditions, the particle-mesh Ewald (PME)~\cite{darden_particle_1993,essmann_smooth_1995}  should be used instead.  For more discussion about the importance of electrostatic interactions, please see Ref.~\citenum{cisneros_classical_2014}. 
The cut-off distance for the short-range neighbor list was \unit[1.2]{nm} and the neighbor lists were  updated at every  step. The intention was to evaluate the ground state energies of DPE molecules with different torsional angle between aromatic rings. To do that, after the first energy minimization step, a short MD run at very low temperature was used to bring the system out of possible local minima. Then a second conjugate gradient energy minimization was performed to obtain the ground state MM/MD PES. 

DFT optimizations and relaxed PES scans were performed using the B3LYP exchange correlation functional \cite{becke_density-functional_1993,lee_development_1988,vosko_accurate_1980,stephens_ab_1994} and def2-TZVP basis set{\cite{weigend_balanced_2005} as implemented in the Orca package~\cite{neese_orca_2012}. Due to the lack of van der Waals (dispersion) interactions in standard DFT, Grimme's DFT-D3 method~\cite{grimme_semiempirical_2006}, was employed.

In order to calculate electronically excited states, many-body Green's function theory in the $GW$ approximation with the Bethe-Salpeter equation (\textit{GW}-BSE)~\cite{hedin_effects_1969} was employed, since static DFT~\cite{kohn_self-consistent_1965} cannot describe coupled electron-hole excitations. For details of the application to molecular systems, the reader is referred to Refs.~\citenum{ma_modeling_2010,baumeier_excited_2012,baumeier_frenkel_2012,onida_electronic_2002,blase_charge-transfer_2011,faber_electronphonon_2012}. The \textit{GW}-BSE method is based on a set of Green's function equations of motion which contain electron-hole interaction (BSE) leading to the formation of excitons. It utilizes the DFT molecular orbitals and energies to calculate the one-particle Green's function ($G$) and screened Coulomb interaction ($W$) to obtain single-particle excitations within the \textit{GW} approximation as introduced by Hedin and Lundqvist~\cite{hedin_effects_1969}. 
An electron-hole excitation cannot be described in an effective single-particle picture but instead requires explicit treatment of a coupled two-particle system. 
The electron-hole amplitudes and associated transition energies can be obtained by solving the Bethe-Salpeter equation~\cite{baumeier_excited_2012,baumeier_frenkel_2012,onida_electronic_2002}. 
For calculation of excitation energies according to \textit{GW}-BSE method, first DFT calculations were performed using the Orca package \cite{neese_orca_2012}, B3LYP functional~\cite{becke_density-functional_1993,lee_development_1988,vosko_accurate_1980,stephens_ab_1994}, effective core potentials of the Stuttgart/Dresden type~\cite{bergner_ab_1993}, and the associated basis sets that are augmented by additional polarization functions~\cite{krishnan_self-consistent_1980} of $d$ symmetry. The specialized $GW$-BSE implementation for isolated systems~\cite{ma_excited_2009,ma_modeling_2010,baumeier_excited_2012,baumeier_frenkel_2012} available in the VOTCA software package~\cite{ruhle_microscopic_2011,note_VOTCA} is used in all further steps related to the excitations. For molecular visualizations, Visual Molecular Dynamics (VMD)~\cite{humphrey_vmd:_1996} and Jmol~\cite{jmol} were used.

\section{Results}
\label{sec:results}

\subsection{Force field parametrization}
Due to the influence of conformational details on the optical properties~\cite{bunz_polyaryleneethynylenes_2009,miteva_interplay_2000} in PPEs, one needs to determine if the force field yields reliable minimum energy configurations.
Hence, relaxed scans  of potential energy surface (PES) were obtained using both MM/MD and DFT.

\begin{figure}
\includegraphics[width=\linewidth]{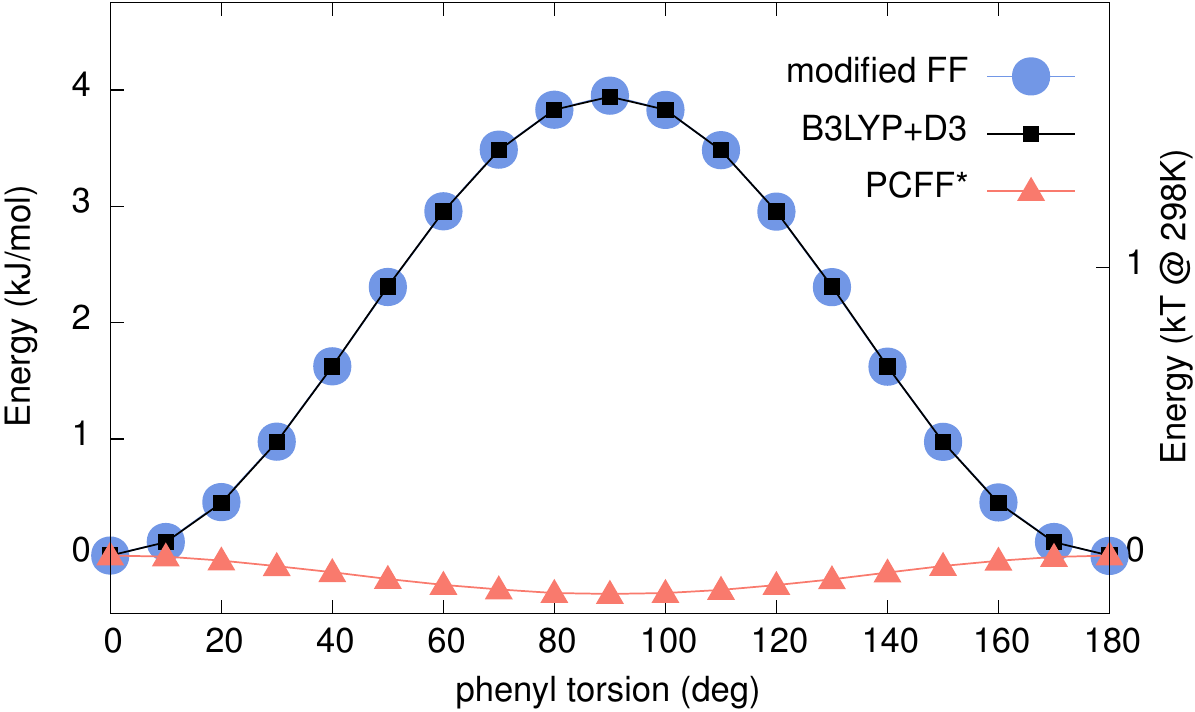}
\caption{Potential energy surface (PES) obtained by MM/MD and quantum mechanical (QM) calculations. Red triangles show PES calculated using $\textnormal{PCCF}^{*}$ \cite{maskey_conformational_2011,maskey_internal_2013} force field. Black squares are the QM results using B3LYP+D3. Blue circles show PES obtained using our new modified force field. The modified force field and B3LYP+D3 are in excellent agreement.}
\label{DPE-QM-MM-PES-vs-torsional-angle}     
\end{figure}

The resulting PES are shown in Fig.~\ref{DPE-QM-MM-PES-vs-torsional-angle}. The PCFF* result (red triangles) shows a minimum at $90^{\circ}$, corresponding to twisted phenylene rings. In contrast, the result of the DFT-based scan (black squares) indicates a minimum energy configuration in which the two phenyl rings are co-planar, which is also extracted from experiments~\cite{stang_acetylenes_1995,halkyard_evidence_1998}. The force field predicts a practically free rotation of phenylenes for $T\geq 0$, while a barrier of around $\sim \unit[4]{kJ/mol}$ $(\sim \unit[1.5]{k_{\text{B}}T})$ is obtained with DFT. The latter is comparable to the one reported in Ref.~\citenum{toyota_rotational_2010}. The experimental potential barrier is around $\sim \unit[2]{kJ/mol}$~\cite{stang_acetylenes_1995,halkyard_evidence_1998,okuyama_electronic_1984}. Overall, the scans imply that the PCFF* force field does not correctly model the  ground state conformations of DPE, which can have severe implications for the derived optical properties. 

\begin{figure}
\centering
\resizebox{1.0\columnwidth}{!}{
\includegraphics{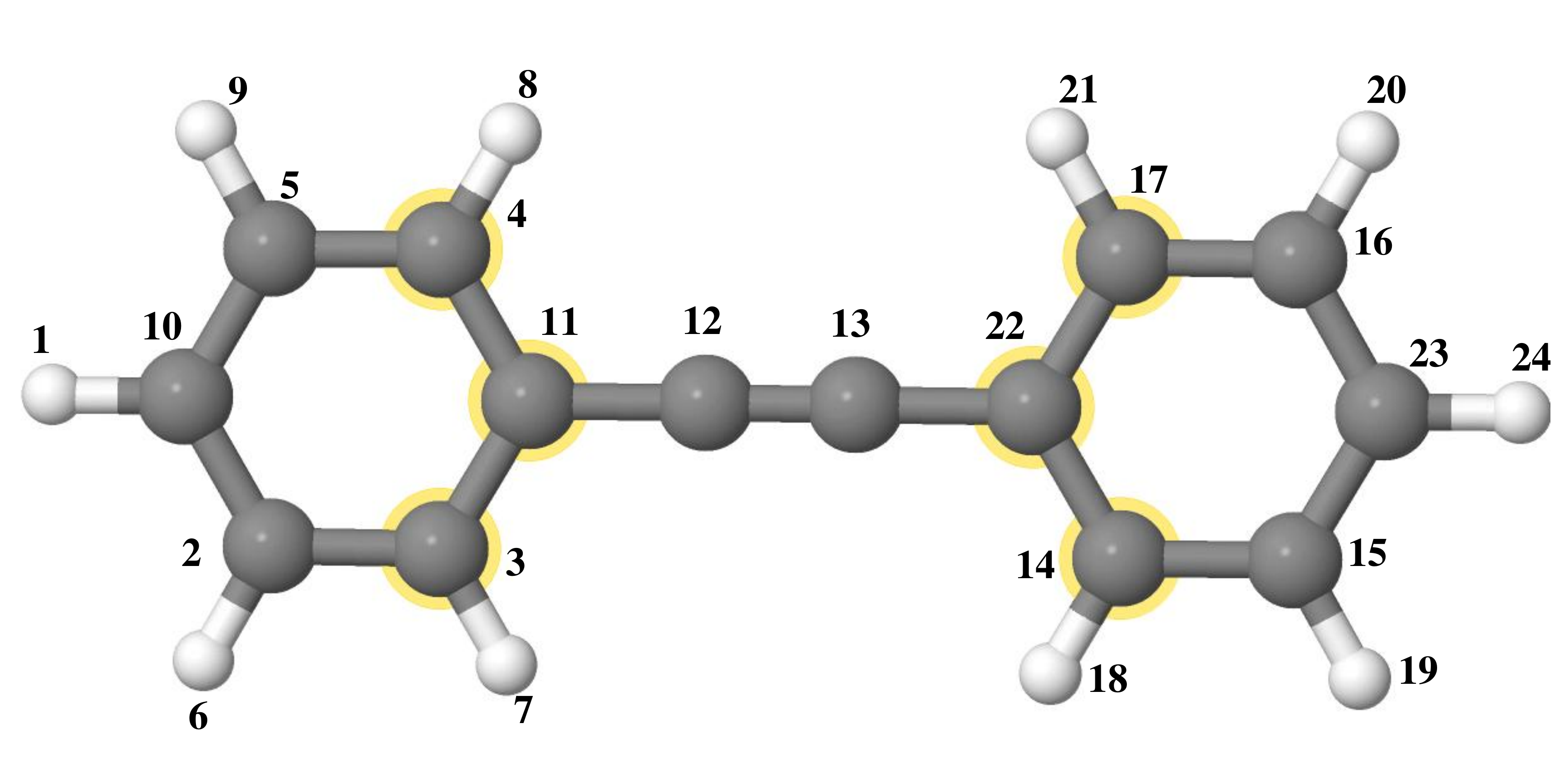}}
\caption{Atomic structure of DPE.  Gray spheres show carbon atoms and white spheres indicate hydrogen atoms. The indices show atom number. There is a triple bond between 12 and 13. }
\label{DPE-chem-number}     
\end{figure}
\begin{table}

\caption{Ryckaert-Bellemans~\cite{ryckaert_molecular_1978} torsion parameters (Eq.~\ref{torsion}) in \unit[]{kJ/mol} for atom numbers 4-11-22-17, 3-11-22-14, 4-11-22-14 and 3-11-22-17, see Fig.~\ref{DPE-chem-number} for  the definition of atom numbers.}
\label{tab:TorsionParameters}       
\resizebox{1.0\columnwidth}{!}{
\begin{tabular}{lccccccccccccc}
\hline\noalign{\smallskip}
 Torsion Type & $k_0$ & $k_1$ & $k_2$ & $k_3$ & $k_4$ & $k_5$\\
\noalign{\smallskip}\hline\noalign{\smallskip}
 C-C-C-C &1.0685&  0.0007&  -1.0660&  0.00004& -0.00375&  -0.0004&
\\
\noalign{\smallskip}\hline
\end{tabular} }
\end{table}

To remedy this situation, the existing force field is refined by the addition of a torsional potential between the two adjacent phenylenes (see Fig.~\ref{DPE-chem-number} for definition of involved atoms). By fitting Eq.~\ref{torsion} to the differences of DFT and PCFF* potential energy surfaces, corresponding Ryckaert-Bellemans~\cite{ryckaert_molecular_1978} force parameters, provided in Table~\ref{tab:TorsionParameters}, were obtained. The PES is re-calculated with MM/MD using the modified force field, yielding the scan as shown in Fig.~\ref{DPE-QM-MM-PES-vs-torsional-angle} (blue circles). It is in good agreement with the DFT result.

\begin{figure}
\includegraphics[width=\linewidth]{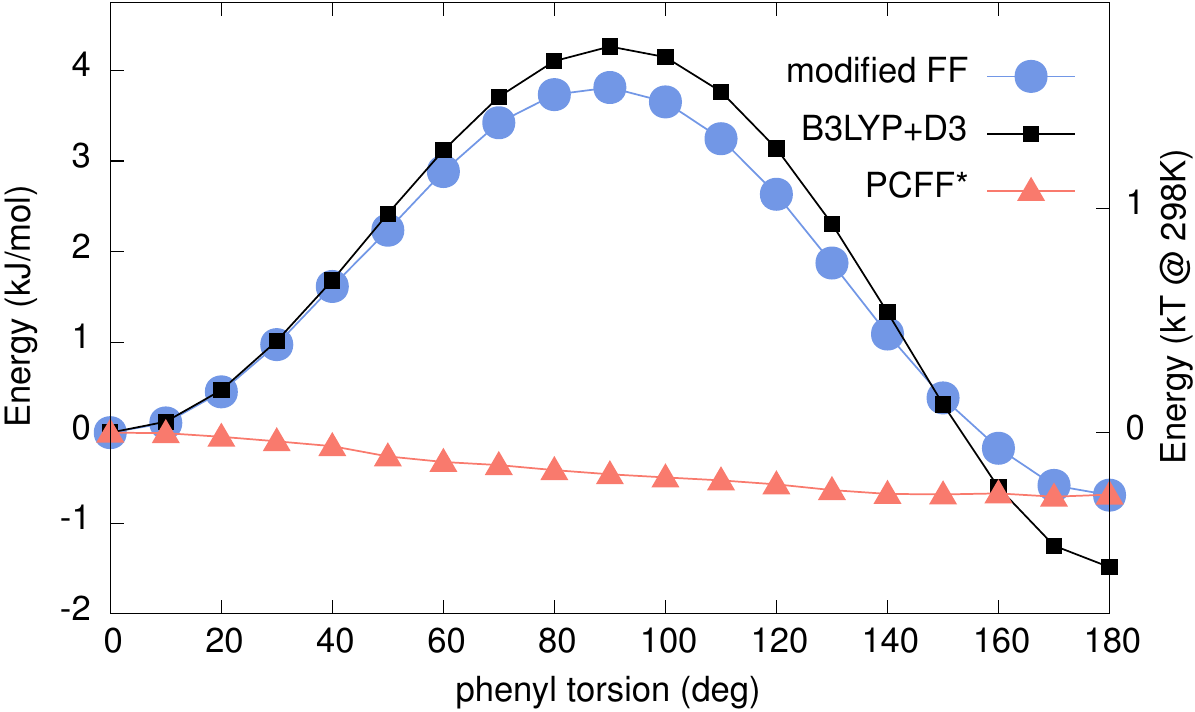}
\caption{Potential energy surface (PES) obtained using the modified force field (black squares), DFT calculations (blue circles), PCFF* force field (red triangles). The modified force filed gives reasonable agreement with the B3LYP+D3. Dispersion interaction between the methyl side chains leads to \textit{cis} conformation (180$^{\circ}$) preference over \textit{trans} conformation (0$^{\circ}$).  }
\label{DPE-methyl-QM-MM-PES-vs-torsional-angle}  
\end{figure}

To assess the transferability of the modified force field, we repeat the above scans of the torsional potential for {\em para} methylated-DPE (see chemical structure shown in Fig.~\ref{DPEstructure}). The PES resulting from both MD and DFT calculations are shown in Fig.~\ref{DPE-methyl-QM-MM-PES-vs-torsional-angle}. With the modified force field (blue circles) one can observe a good agreement with the DFT data (black squares). Both approaches predict a minimum energy configuration at $180^{\circ}$ twist. The energetic preference of this \textit{cis} conformation of Me-DPE over the \textit{trans} conformation (0$^{\circ}$) is driven by attractive dispersion interaction among the two CH$_3$. While this preference is also obtained with the original PCFF$^*$ force field (red triangles), no barrier between \textit{cis} and \textit{trans} configurations is found.
In terms of obtaining minimum energy configurations and energy barriers in the PES, the modified PCFF is clearly more reliable.

\subsection{Optical excitations in single molecules}
For systems such as solvated polymer chains, the system size makes the use of classical simulations inevitable to obtain structural information. Even with the modified force field at hand, it is not automatically guaranteed that the use of the MM/MD geometries in QM/MM schemes does not lead to spurious errors in the computed excitations. To further assess the level of reliability of such calculations, the evolution of optical absorption properties of Me-DPE is examined as a function of phenyl torsions based on the respectively optimized geometries. 

\begin{figure}
\includegraphics[width=\linewidth]{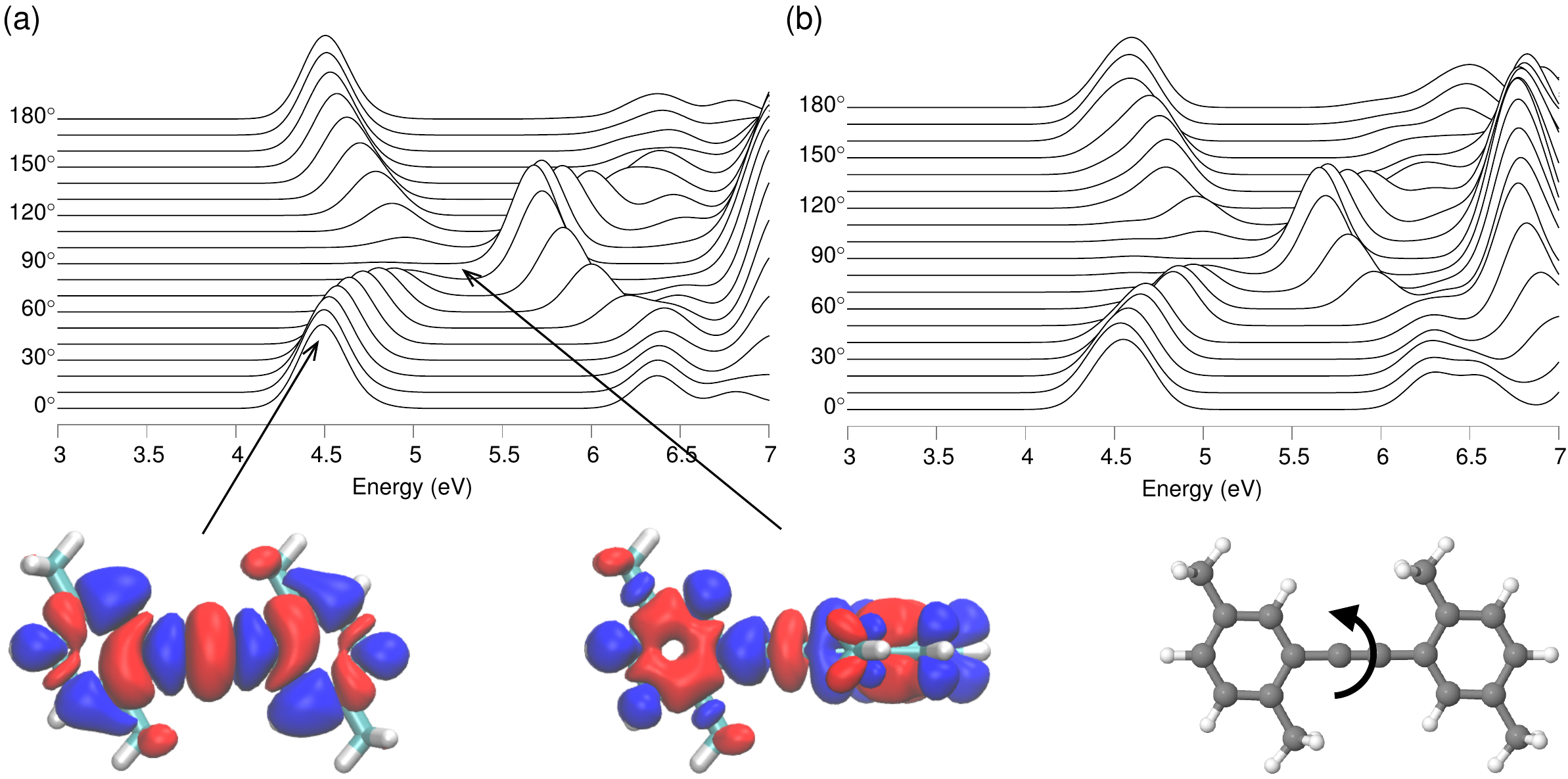}
\caption{Top: Optical absorption spectra for Me-DPE as a function of torsional angle between phenylene rings based on (a) DFT optimized structures and (b) MM/MD energy minimized structures. The lower energy excitations in both (a) and (b) show the same dependency on the angle. Bottom: isosurfaces ($\unit[\pm5\times 10^{-3}]{e/\text{\AA}^3}$) of excitation electron density at  $0^{\circ}$ and $90^{\circ}$ based on DFT optimized structures. Red color corresponds to negative values (hole density) and blue color corresponds to positive values (electron density). Electron and hole densities are extended along the molecule at $0^{\circ}$ and $90^{\circ}$ and no localization of the excitation occurs. }
\label{fig:waterfalls}
\end{figure}

The optical absorption spectra resulting from $GW$-BSE are shown in Fig.~\ref{fig:waterfalls}. DFT optimized geometries were used in (a) and MD energy minimized geometries using the modified force field were used in (b). The height of the curves indicates the strength of the excitation. Comparing both results, it is evident that  the same dependency on the torsional angle is obtained by both approaches. With increasing twist from $\unit[0]{^\circ}$ to $\unit[90]{^\circ}$ the main absorption peak gradually shifts to higher energies while its strength decreases at the same time until it vanishes at $\unit[90]{^\circ}$. Inspection of the electron and hole densities of the excitations for co-planar and perpendicular (see bottom of Fig.~\ref{fig:waterfalls}) also reveals no localization of the excitation during the rotation confirming that the conjugation via the C$\equiv$C bond is indeed strong. The identical behavior of the lowest energy excitations (which are typically those of interest) for both MM/MD and DFT conformations indicates that the modified force field is suitable for use in QM/MM calculations. 

\begin{figure}
\includegraphics[width=\linewidth]{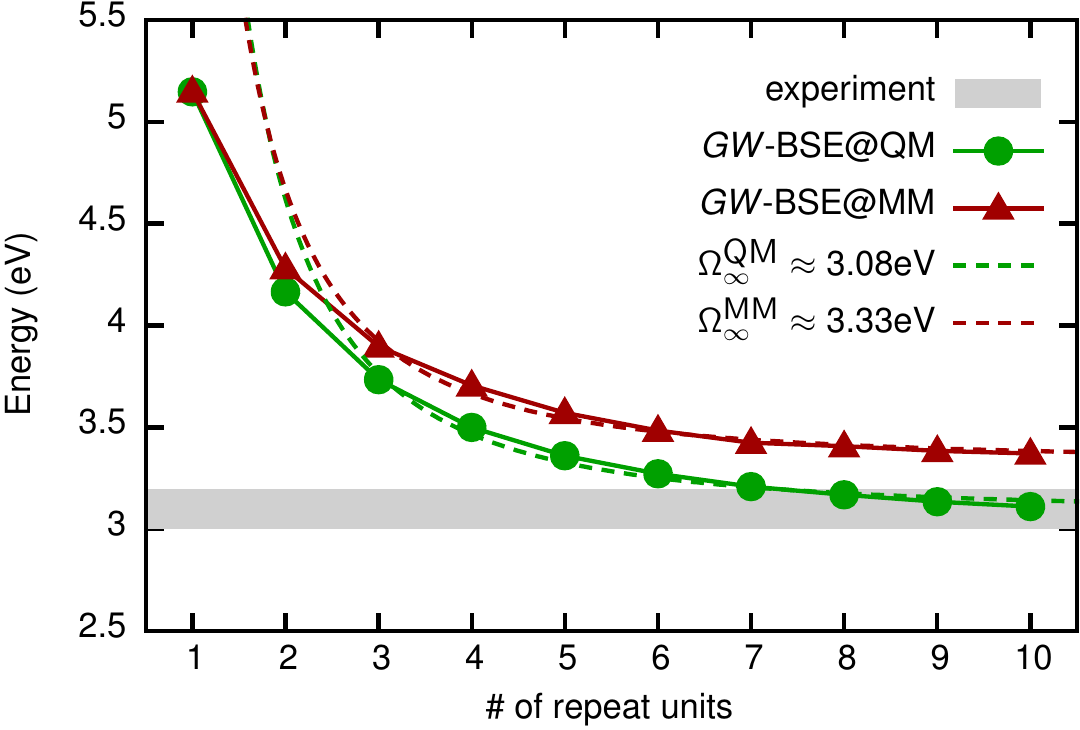}
\caption{Lowest optically active excitation energies in $n$-PPE as the number of repeat units is increased in from $n$=1 to $n$=10. Results obtained on DFT (MM/MD) geometries are shown as green points (red triangles). the respective dashed lines indicate the fit to the quantum-size model. The gray shaded area indicates the width of the experimental data~\cite{halkyard_evidence_1998}.}
\label{fig:optical-oligomers}
\end{figure}

So far the analysis has been limited to small model systems. As a next step towards more realistic system sizes, {\em para} phenylene ethynylene (PPE) oligomers (see Fig.~\ref{PPE-chem}) are investigated. Increasing the number of repeat units $n$ from one to 10, lowest optically active excitation energies were computed with $GW$-BSE for geometries optimized using quantum (\textit{GW}-BSE@QM) and force field (\textit{GW}-BSE@MM) approaches, respectively. The results shown in Fig.~\ref{fig:optical-oligomers} exhibit a monotonous decrease with $n$ for both approaches. Such a strong size-dependence can be traced back to an increase in the size of the conjugated system. From the particle-in-a-box model, one can estimate, e.g., the optical excitation energy of an infinitely long chain via $\Omega(n) = \Omega_\infty - a/n$. By fitting the data for $n>3$ to this model, a value of $\Omega^\text{QM}_\infty = \unit[3.08]{eV}$ is obtained for QM geometries. For  
$n>7$ the respective excitation energies vary only slightly and approach the region in which experimental absorption is measured in experiment~\cite{halkyard_evidence_1998}. This indicates that studying more complex morphologies, i.e., solvated polymers, based on oligomers with $n=10$ is an adequate choice. For MM geometries, the absorption energies result slightly higher, evidenced by the estimate of $\Omega^\text{MM}_\infty = \unit[3.33]{eV}$. Upon further inspection, this offset of \unit[0.25]{eV} with respect to $\Omega^\text{QM}_\infty$ is a cumulative result of slight discrepancies in bond length within the phenylenes and the $C-C$ bridge bonds. In conclusion, the use of geometries determined using MM/MD in $GW$-BSE calculations can be expected to lead to slight quantitative overestimates of excitation energies. Qualitatively, however, a satisfying agreement is found.

\subsection{Absorption and emission of solvated 2,5-dinonyl-10-PPE}
\begin{figure}
\includegraphics[width=\linewidth]{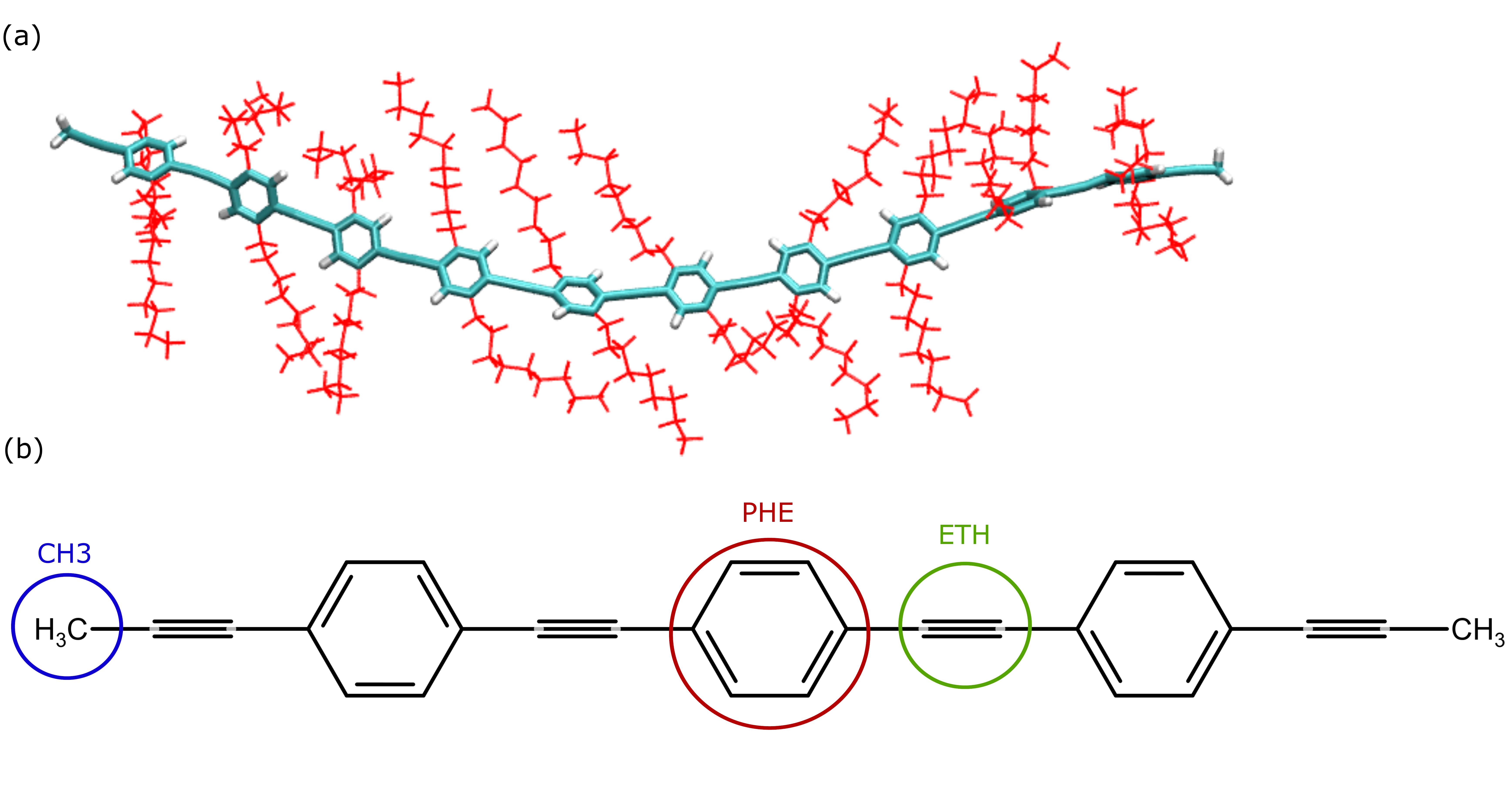}
\caption{(a) Sample configuration of 2,5-dinonyl-10-PPE solvated in toluene. Nonyl side chains are indicated in red and solvent molecules are not shown for clarity. (b) Definition of three types of rigid fragments used in back mapping of the backbone conformations used in the QM/MM setup.}
\label{fig:fragments}
\end{figure}

As an example of a typical application of a combined quantum-classical simulation of optical excitations, a single chain of 10-PPE was functionalized by nonyl side chains in 2,5 positions of the phenyl rings and solvated by toluene. The modified PCFF force field for the PPE backbone is used in combination with OPLS for nonyl and toluene. For details of the simulations, see Ref.~\citenum{bagheri_solvent_2016}. A sample configuration of backbone and side chains is shown in~\Fig{fragments}(a). Since toluene is a poor solvent for the nonyl, one can observe extended and partially strongly interacting side chains. As a consequence, the backbone is under considerable non-uniform stress leading to the overall curvature of the usually rigid polymer. 

A set of 11 snapshots with a time step of \unit[1]{ps} is taken from the classical MD trajectory and each of the snapshots is partitioned into a quantum (the backbone) and a classical region comprising the side chains and the solvent molecules. QM and MM regions interact via static partial charge distributions. The aim of this setup is to evaluate the excitations of the polymer backbone taking its curved conformation into account while reducing discrepancies between the force-field and QM geometries, as much as possible. At the same time, the bridging carbon-carbon bond between the phenyl 2 and 5 positions and the nonyl side chain, defining the boundary between QM and MM regions, needs to be broken and the dangling bond saturated by hydrogen atom. This can be achieved with the help of a re-mapping scheme based on the definition of molecular fragments. Using centers of mass and gyration tensors, fragments of optimized QM configurations were mapped onto the orientation and alignment of the corresponding fragments in the MD configurations. 

\Fig{fragments}(b) illustrates the re-mapping scheme for PPE. Each phenyl ring (PHE), ethyne pair (ETH), and terminal methyl group (CH3) is defined as a unique fragment. A 10-PPE backbone is hence subdivided into a total of 23 fragments (10 PHE, 11 ETH, 2 CH3) for mapping purposes. 

\begin{figure}
\includegraphics[width=\linewidth]{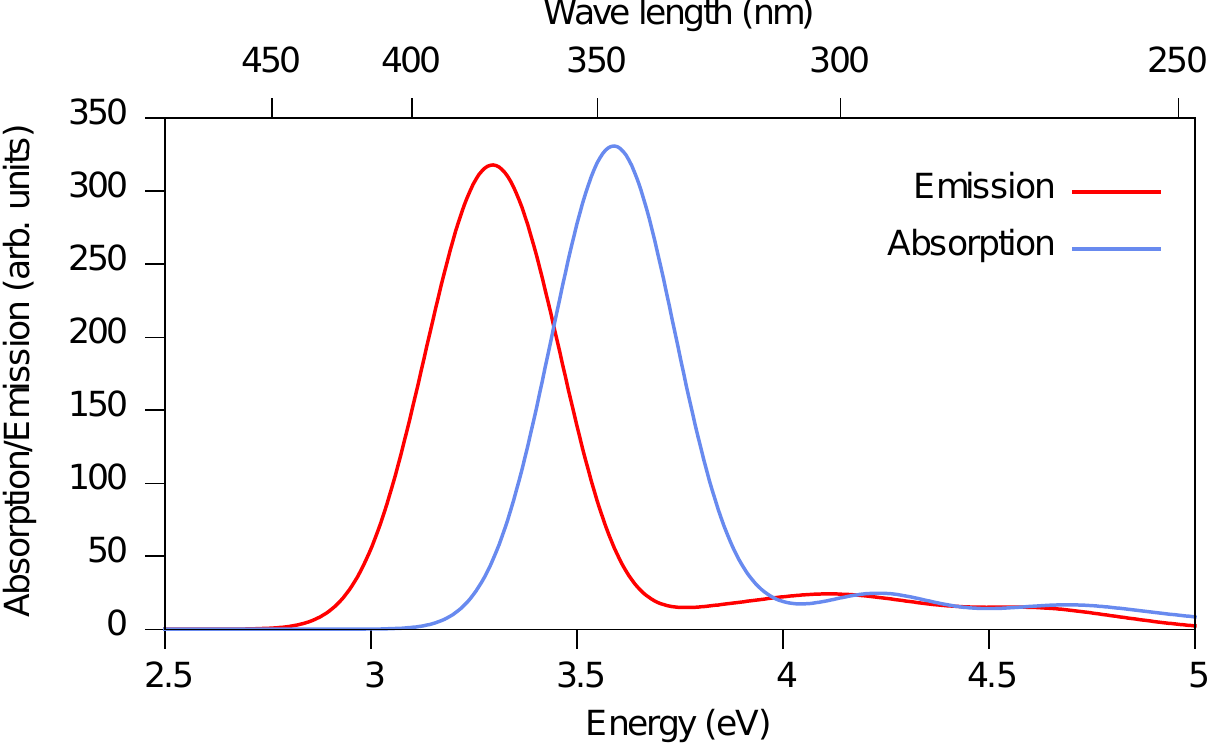}
\caption{Optical absorption (blue) and emission (blue) spectra for 2,5-dinonyl-10-PPE in toluene as obtained from $GW$-BSE/MM calculations using fragment based re-mapping. Both spectra are averages over 11 snapshots taken every \unit[1]{ps}, respectively. A Stokes shift (red shift) of \unit[0.29]{eV} is found, which compares well to experimental observations of \unit[0.3-0.4]{eV}~\cite{halkyard_evidence_1998}.}
\label{fig:emission}
\end{figure}

With the re-mapped conformations at hand, the coupled $GW$-BSE/MM system is solved and the absorption spectrum determined as an average over the eleven snapshots. Individual spectra are broadened by Gaussian functions with a FWHM of \unit[0.3]{eV} and the resulting spectrum is shown as a blue line in \Fig{emission}. It is characterized by a single peak at an energy of \unit[3.64]{eV}, which is larger than the value of \unit[3.11]{eV} obtained for an isolated single oligomer. This spectral blue shift is a direct result of the polymer's curvature constrained by the side chain interactions. With the re-mapping scheme it is also possible to approximate emission spectra by using excited state QM geometries as a reference~\cite{note_ESGeo}. Upon excitation, electrons are promoted to higher, often anti-bonding, molecular orbitals causing an extension of bonds. Constrained by side chains, a more general modification of the overall conformation can (at least on short time scales) not be expected. Solving the $GW$-BSE/MM system based with excited state re-mapping yields the emission spectrum shown as a red line in \Fig{emission}. While the no changes in the spectral shape can be noted, the peak position of the emission is red-shifted by \unit[0.29]{eV} compared to the absorption peak. This Stokes shift is in good agreement with experimental data in the range of \unit[0.3-0.4]{eV}~\cite{halkyard_evidence_1998}.

\section{Conclusions}
\label{sec:conclusions}
A combination of atomistic (MM/MD) and DFT calculations were performed to describe conformational  properties of diphenylethyne (DPE), methylated-DPE and poly para phenylene ethynylene (PPE). MM/MD simulations based on PCCF* force field were not able to provide a good description of the ground state conformation of the DPE molecule. Due to this, DFT calculations were employed to develop force field parameters to improve the MM/MD simulations. The modified force field was able to describe the conformation  of methylated-DPE in  agreement with DFT results. The $GW$-BSE method was utilized to describe excited states of the methylated-DPE and $n$-PPE polymer with $n=1,2,\dots,10$. Optical excitations were obtained for the methylated-DPE and $n$PPE based on MM/MD energy  minimized structures using the modified force field and DFT optimized geometries. The results for methylated-DPE  show that the lowest energy excitations based on the MM/MD conformations and DFT optimized geometries  follow the same pattern. This nearly identical behavior for the lowest energy excitations indicates that one can describe optical excitations using the $GW$-BSE method based on  MM/MD conformations. Results for the excitation energies for $n$PPE indicate that there is an overall agreement between the results of $GW$-BSE based on MM/MD energy minimized structures and DFT optimized geometries. There is a discrepancy of around 0.25~eV between the two. This discrepancy is a cumulative result of geometric differences between MM/MD and DFT structures. Overall agreement between MM/MD  and QM based excitations  is enough to validate the use MM/MD conformations as the basis for calculation of optical excitations with $GW$-BSE method.


\begin{mcitethebibliography}{91}
\bibitem[not()]{note_ESGeo}
The lack of analytical forces for $GW$-BSE makes the optimization of the
  excited state geometries of molecules of the size of 10-PPE practically
  impossible. Therefore, the geometry of the first excited state is optimized
  using time-dependent DFT (TDDFT), with the same functional and basis set as
  in the ground state. Numerical gradients of DFT/$GW$-BSE total energies basd
  on the TDDFT optimized sturcture yields very small forces.\relax
\mciteBstWouldAddEndPunctfalse
\mciteSetBstMidEndSepPunct{\mcitedefaultmidpunct}
{}{\mcitedefaultseppunct}\relax
\EndOfBibitem
\end{mcitethebibliography}

\providecommand*{\mcitethebibliography}{\thebibliography}
\csname @ifundefined\endcsname{endmcitethebibliography}
{\let\endmcitethebibliography\endthebibliography}{}

\end{document}